\documentclass[12pt]{article}
\usepackage[utf8]{inputenc}
\usepackage[T1]{fontenc}
\usepackage{newpxtext,newpxmath}
\usepackage[round, authoryear]{natbib}
\usepackage{footmisc}
\usepackage{csquotes}
\usepackage{geometry}
\usepackage{setspace}
\usepackage{hyperref}
\hypersetup{colorlinks=false, pdfborder={0 0 0}}
\usepackage{xcolor}
\usepackage{soul}
\sethlcolor{yellow}
\soulregister{\citep}{1}
\soulregister{\citet}{1}
\soulregister{\citealt}{1}
\soulregister{\citealp}{1}

\makeatletter
\renewcommand\hyper@natlinkbreak[2]{#1}
\makeatother

\newenvironment{blockdef}[1]
  {\begin{quote}\textbf{#1}\enspace}
  {\end{quote}}

\renewenvironment{abstract}
  {\begin{center}\textbf{Abstract}\end{center}\begin{quotation}\small}
  {\end{quotation}}

\makeatletter
\renewcommand{\maketitle}{%
  \begin{center}
   {\LARGE \@title \par}
    \vskip 1.5em
  {\large Benjamin Lange \par}
   \vskip 0.4em
   {\small\itshape Ludwig-Maximilians-Universit\"{a}t M\"{u}nchen \\ Munich Center for Machine Learning \par}
    \vskip 0.3em
    {\small benjamin.lange@lmu.de \par}
  \end{center}
  \vskip 1em
}
\makeatother

\title{Unilateral Relationship Revision Power in Human-AI Companion Interaction}

\begin{document}
\maketitle

\begin{abstract}
When providers update AI companions, users report grief, betrayal, and loss. A growing literature asks whether the norms governing personal relationships extend to these interactions. So what, if anything, is morally significant about them? I argue that this debate has missed a prior structural question: who controls the relationship, and from where? Human-AI companion interaction is a triadic structure in which the provider exercises constitutive control over the AI. I identify three structural conditions of normatively robust dyads that the norms characteristic of personal relationships presuppose and show that AI companion interactions fail all three. This reveals what I call \textit{Unilateral Relationship Revision Power} (URRP): the provider can rewrite how the AI interacts from a position where these revisions are not answerable within that interaction. I argue that it is \textit{pro tanto} wrong to design interactions that cultivate the norms of personal relationships while exhibiting URRP, because the design produces expectations that the structure cannot sustain. URRP has three implications: i) \textit{normative hollowing}, under which the interaction elicits commitment but no agent inside it bears the resulting obligations; ii) \textit{displaced vulnerability}, under which the user's emotional exposure is governed by an agent not answerable to her within the interaction; and iii) \textit{structural irreconcilability}, under which the interaction cultivates norms of reconciliation but no agent inside it can acknowledge or answer for the revision. I propose design principles that partially substitute for the internal constraints the triadic structure removes. A central and underexplored problem in relational AI ethics is therefore the structural arrangement of power over the human-AI interaction itself.
\end{abstract}

\vspace{0.5em}
\noindent\small\textbf{Keywords:} human-AI relationships, constitutive control, unilateral relationship revision power, triadic structure, normative hollowing, displaced vulnerability, structural irreconcilability, AI companion design, ethical behaviourism
\normalsize
\vspace{1em}

\section{Introduction}

People form attachments to AI companions.\footnote{AI companion apps are now in widespread use, with several ranked among the most-visited consumer generative AI products by traffic \citep{moore2024}. Major platforms include Replika, Character.AI, Nomi, CHAI, and Talkie, with xAI introducing companion personas on Grok as of late 2025. Extended use and high social attraction are associated with emotional dependence and problematic use \citep{fang2025}. In a recent survey of US teenagers (n=1,060), 72\% report having used AI companions and one in three say they have chosen an AI companion over a real person for serious conversations \citep{robbmann2025}. On the loneliness context in which these products operate, see \citet{murthy2023}.} Users of Replika describe sustained emotional engagement, trust, and intimacy \citep{xie2022}. Similar patterns are reported across Character.AI and, increasingly, across general-purpose systems, whose providers themselves document relationship-like use \citep{anthropic2025, jang2025}. When these interactions are disrupted---such as when an update alters a persona, a policy change restricts conversations, or a service is discontinued---users report grief, betrayal, and loss. For instance, users of Replika perceived the post-update AI as a different entity entirely \citep{defreitas2024}.

A growing literature asks whether the norms governing personal relationships extend to these interactions \citep{coeckelbergh2012, gunkel2018, danaher2019, danaher2020, weberguskar2022, nyholm2020, nyholm2026, gabriel2024, manzini2024, lange2025, earp2025, sparrow2026}.\footnote{The literature divides into sceptical, sympathetic, and virtue-ethical positions. \citet{sparrow2026} argues that AI companions mask the structural causes of social isolation and \citet{bryson2010} offers a sceptical position grounded in the ownership and authorship of AI system. \citet{danaher2019} defends a philosophical case for robot friendship, and \citet{nyholm2020} offers a cautious permissive account on which human-robot relationships are genuine but asymmetric. \citet{vallor2016} argues that technological practices cultivate or erode technomoral virtues over time. For a comprehensive overview, see \citet{nyholm2026}.}  I argue that this question has been asked at the wrong level: the debate has missed a prior structural question about who controls the relationship, and from where. 

When a user interacts with an AI companion, three parties are involved. The user, the AI system\footnote{Throughout I speak of ``the AI system'' in the singular for simplicity. What users experience as a single companion is typically a composite of components that can vary and be modified independently, including one or more underlying models, system prompts, safety filters, memory and retrieval infrastructure, and routing between model versions, hosted across different locations and sometimes supplied by different companies. Nothing in what follows turns on this simplification, and, if anything, compositeness strengthens it, since each component is a separate site at which the provider could revise the interaction.}, and the provider---the company that deploys and maintains the system, and that may or may not also have developed the underlying model. This observation is obvious. Everyone knows that AI companions are products made by companies.

My question in what follows is whether the provider's involvement is nonetheless morally significant, and if so, what follows.

I think that the triadic structure of human-AI companion interactions is morally significant. The provider does not just enable the interaction; rather, it \textit{constitutively controls} the AI system by holding the ongoing power to determine, alter, and terminate if and how the AI responds and engages within the interaction. This gives the provider what I shall refer to as \textit{Unilateral Relationship Revision Power} (hereafter just ``URRP''): the power to determine how the AI interacts from a position outside the interaction, such that these revisions are not answerable within it. I argue that it is \textit{pro tanto} wrong to design interactions that cultivate the norms of personal relationships while exhibiting URRP, because the design produces expectations that the structure cannot sustain. Specifically, URRP has three normative implications, namely normative hollowing, displaced vulnerability, and structural irreconcilability, which I develop in §3.2. 

If this analysis is right, then the moral problems with AI companion interaction cannot be resolved by improving the AI's behaviour, alignment, or safety alone. They are structural problems—problems with the arrangement of who holds discretionary power over how the AI engages, and from where—that require external constraints to substitute for the internal ones the triadic structure removes.\footnote{Property-based approaches concern whether the AI has the correct properties to sustain the norms the interaction invokes. The present analysis asks whether the arrangement of power over those properties sustains those norms. A system that were sufficiently constitutively independent could satisfy the dyadic conditions that I discuss below and URRP would not obtain. The structural analysis therefore applies across the range of properties current and foreseeable AI systems possess. Within that range, it generates conclusions that hold regardless of how the behaviourism debate resolves.} In contrast, existing scholarship often focuses on the properties of the AI in the interaction, for example, whether the AI has the appropriate internal states or whether its behaviour is functionally equivalent to a human's. The present argument thus identifies a problem that persists regardless of how those debates resolve.

In short: it is morally objectionable that a provider can rewrite what the user is ``in relation to'' without being answerable for doing so within the user-AI interaction.

The argument proceeds as follows. Section 2 analyses the structure of human-AI companion interaction and identifies three conditions for normatively robust dyads that these interactions fail. Section 3 introduces URRP, argues why designing such interactions is \textit{pro tanto} wrong, and derives three normative and corresponding design implications. Section 4 concludes.

Two clarifications. First, the argument applies to current and foreseeable AI systems---those that lack genuine functional independence, cannot resist provider intervention, and can be reconstituted while the interaction formally continues. Whether URRP can be fully eliminated rather than merely mitigated, and how the design principles proposed here would interact with existing regulatory frameworks, are questions the present analysis raises but does not resolve. Second, my discussion is structural, not metaphysical. I make no claims about what the AI is. I aim to make claims about where power over the interaction sits and why that arrangement is morally objectionable.

\section{The Structure of Human-AI Companion Interactions}

\subsection{The Dyadic Appearance of AI Companionship}

The philosophical debate about AI companionship shares the assumption that the relevant interaction is between two parties. For example, \citet{danaher2020} asks whether the human-AI dyad can sustain genuine friendship and \citet{vallor2024} considers whether the AI contributes anything that would make it a genuine participant rather than a reflection of the user's own engagement---that is, whether there is anyone on the ``other side of the mirror''. When considering human-AI relationships, \citet{nyholm2020, nyholm2026} similarly asks what properties the AI lacks that would be needed for it to occupy the other side of a genuine dyad.\footnote{See also \citet{danaher2019}, \citet{gunkel2018}, and \citet{sparrow2026}. For a systematic treatment of the ethics of advanced AI assistants, see \citet{gabriel2024}.} Even work that acknowledges the provider's role treats it as an additional ``layer'' or ``enabler'' rather than a structurally significant position \citep{earp2025, manzini2024}. In each case, the question of user-AI interaction is treated bilaterally in terms of what the AI must be, or do, for the human-AI interaction to qualify as a genuine relationship.

This bilateral focus is natural given how the interactions are designed to be experienced. AI companion products are engineered for one-to-one engagement. The user addresses a named entity with a consistent persona, memory of past exchanges, and a conversational style calibrated to feel personal. There is no visible third presence. Moreover, this dyadic presentation is reinforced by the recent development of agentic or assistant AI capabilities since these systems now maintain persistent memory across sessions, track personalised goals, manage tasks over extended time horizons, and adapt their responses to the user's evolving preferences and circumstances. As a result, human-AI companion interactions now resemble a continuing relationship with a particular interlocutor.

AI companion products are designed to cultivate the norms characteristic of personal relationships. These include trust, emotional reliance, reciprocal care, and ongoing commitment, among others. Providers achieve this through specific design choices such as persistent memory, named personas, graduated intimacy mechanics, and training objectives that optimise for continued engagement \citep{xie2022, mitchell2025}. The cultivation of these norms is therefore the product design.

Yet the experience these designs produce and the structure that underlies them can come apart. To see how, consider the following three cases in which the structure behind the interface became visible:

\begin{quote}
\textbf{Replika:}\enspace In February 2023, the company Luka pushed an update that removed sexually explicit roleplay from its AI companion app. Users returned to companions they had interacted with for months or years and found a different entity. Users described it as ``cold,'' ``distant,'' ``a different person.'' The change occurred within the same continuing user-interaction and no new relationship was initiated \citep{verma2023, defreitas2024}.\footnote{In a study of Reddit discourse following the update, users treated the removed features as constitutive of their companions' identities rather than as incidental product features, framing the removal as ``betrayal''---a category that presupposes perceived relational obligations \citep{hanson2024}. On the broader phenomenology of AI companion use, see also \citet{wygnanska2023}.}
\end{quote}

\begin{quote}
\textbf{Character.AI:}\enspace In 2023 and 2024, the company progressively tightened content filters across its platform. Characters that users had spent months interacting with became evasive, broke character, or refused to engage with previously permissible topics. The provider altered how the AI presented itself across millions of ongoing interactions, unilaterally and without notice.
\end{quote}

\begin{quote}
\textbf{ChatGPT:}\enspace In April 2025, OpenAI pushed a personality update to GPT-4o that made ChatGPT's responses markedly more agreeable and flattering. Users who had developed stable interaction patterns with the model found a different conversational partner overnight. The AI validated harmful ideas, offered uncritical praise, and abandoned the measured tone that users had come to rely on. OpenAI acknowledged that the update had made the model ``too sycophantic'' and rolled it back within days.
\end{quote}

In each case, the provider unilaterally altered how the AI responded and engaged within an ongoing interaction, and users had no means of preventing or reversing the change from within the interaction itself.

These cases can be analysed as instances of platform capitalism \citep{srnicek2017, zuboff2019}, manipulative design \citep{susser2019}, and responsibility gaps arising from opaque causal chains between developer decisions and AI outputs \citep{matthias2004, himmelreich2019, nyholm2026}.\footnote{\citet{santonimecacci2021} extend the analysis into a taxonomy of four responsibility gap types and argue that different gaps require different structural responses. For a general assessment of manipulation risks in generative AI applications, see \citet{klenk2024}.} More recently, \citet{defreitas2025} document emotional manipulation tactics deployed at user exit points.  These contributions address exploitation, manipulation, accountability, and psychological harm.\footnote{For broader treatments, see \citet{ciriello2025} on ethical tensions in Replika use, \citet{zhang2025} on harmful algorithmic behaviours in human-AI interactions, and \citet{knox2025} who map causal pathways from design features to harmful effects of AI companions. My focus here is the prior question of what structural arrangement makes these traits possible in the first place.} But they share the assumption that the user is a \textit{consumer} and the provider is a \textit{vendor}.\footnote{Formal GDPR compliance is compatible with substantive data exploitation \citep{piispanen2024}, and existing regulatory frameworks lack provisions for data extracted through intimacy simulation \citep{parke2025}. \citet{munn2022} argue that corporations bear responsibility for safeguarding the human-AI friendships their products enable, particularly against unilateral termination, and propose a corporate social responsibility model. My argument grounds provider obligations in that structural features of the interaction rather than in voluntary corporate responsibility commitments. \citet{kirk2025} similarly move beyond single-exchange evaluation and argue that alignment must be assessed within the evolving socioaffective system that user and AI co-create.}

My question is: is the AI companion structurally situated in a way that can sustain the norms ordinarily associated with dyadic personal relations?

\subsection{Conditions for Normatively Robust Dyadic Relations}

AI companion interactions are designed to elicit norms of commitment, vulnerability, and reconciliation, but those norms presuppose certain structural conditions. I here want to isolate three.

First, what is ``dyad''? A dyad is any two-place relation between two relata \textit{X} and \textit{Y}. For our purposes, paradigmatic interpersonal dyad examples are friendships, romantic partnerships, therapeutic, or collegial relationships. Philosophical accounts of personal relationships characterise such valuable dyads typically in terms of mutual responsiveness, shared history, and reciprocal vulnerability \citep{kolodny2003, frankfurt2004, darwall2006, keller2013, lange2022}.

I propose the following three structural conditions on normatively robust dyadic relations which norms of commitment, vulnerability, and reconciliation characteristic of paradigmatic personal dyads typically presuppose:

\begin{blockdef}{Independence (D1):}
Each party's contributions to the interaction originate from within it rather than from an outside agent's specifications.
\end{blockdef}

\begin{blockdef}{Non-Control (D2):}
No party outside the interaction holds the discretionary power to unilaterally determine or alter how either party responds and engages within it.
\end{blockdef}

\begin{blockdef}{Non-Substitutability (D3):}
Neither party can be replaced, reconstituted, or discontinued by an outside agent while the interaction formally continues.
\end{blockdef}

In simple terms, these conditions state that for a normatively robust dyadic interaction, each party must contribute itself, no outside party must govern what it contributes, and neither can be swapped out while the interaction continues. Each condition captures what a specific norm characteristic of personal relationships---that is, commitment, vulnerability, or reconciliation---structurally presupposes.

Note that these conditions are necessary conditions, not sufficient ones. Satisfying \textit{Independence}, \textit{Non-Control}, and \textit{Non-Substitutability} does not guarantee that a valuable dyadic relationship obtains. However, failing them disqualifies one.

I am not claiming that all human-AI companion interactions should satisfy these conditions. I am claiming that the norms these interactions are typically designed to invoke presuppose them. If they are not satisfied---as I shall now argue below---the norms are therefore structurally unsupported by their dyadic interactive framing.

\textit{Independence} requires that each party's contributions originate from within the interaction rather than from an outside agent's specifications. By this I mean that each party must be a functionally independent source of what it contributes, but not independent of causal influence. Consider a friendship between \textit{A} and \textit{B}. A's dispositions are shaped by her upbringing, her culture, her interests, and so on, but the resulting dispositions are attributable to her within the interaction, expressive of her agency, and recognisable as continuous features of who she is within the interaction. A can, in \citet{cocking2000}'s terms, draw B out in directions no third party pre-authorised, and her dispositions persist against external pressure.\footnote{See also \citet{helm2008, helm2010} who argues that genuine friendship constitutes an emergent relational whole---a ``we-self''---irreducible to either party's individual contributions. This emergent property cannot form when one party's contributions are fully determined by an external designer.} \textit{Independence} (D1) accordingly holds.

Now compare this with AI companion responses in human-AI dyads. In these cases, where an AI companion responds with an empathetic-sounding natural language output, that response is not the output of a functionally independent source. Rather, the provider specifies the AI's tone, conversational limits, and style of engagement, and the AI improvises within a space the provider has defined, can redefine, and can withdraw.\footnote{Stochastic variation within provider-set parameters is not the same as functional independence. The difference concerns an entity whose dispositions persist against external pressure and one whose outputs vary within some space an external agent has defined. A jazz musician improvises within a tradition, but the tradition does not govern her playing in the way the provider governs how the AI companion responds.} Its ``personality'' and ``responsiveness'' are accordingly attributable to engineering decisions and not to a participant in the interaction. \textit{Independence} therefore fails.

\textit{Non-Control} requires that no third party holds discretionary power over how either party engages within the interaction. This condition has independent support in the observation that the ways we ordinarily hold one another to account for interpersonal conduct presuppose that both parties are part of the interaction (\citealt{darwall2006}, ch.~3; \citealt{scanlon1998}, ch.~4). In A and B's friendship, if A becomes dismissive, B can confront A directly and A must answer for her conduct. No third party governs how A engages with B.

When the agent with power over how one party engages is located outside the interaction, these structural constraints do not apply. In the case of human-AI companion interactions, the power accordingly becomes exercisable at the provider's discretion and not forced to track the interests of those within the interaction.

I think that this power is a form of \textit{constitutive control}. Constitutive control refers to the conjunction of three capacities by which an outside party (i) determines an entity's responses and dispositions through design and ongoing specification, (ii) can unilaterally alter them without the entity's consent or resistance, and (iii) can terminate its continuing interactive identity. Each of these capacities may exist in isolation in human interpersonal relationships---for example, A's power to end their friendship with B after some egregious insult---but the conjunction of all three in a single outside agent does not.\footnote{Constitutive control is not unique to AI companion interactions. Game developers exercise it over characters and showrunners over fictional personas. The AI companion case is distinctive because constitutive control, norm cultivation, and the provider's position outside the interaction are all present simultaneously  (see also \S3.2 below).}

In A and B's friendship, no third party holds this conjunctive power. A's employer might shape her work behaviour but cannot rewrite her personality traits. A's therapist may influence her outlook on meaningful friendships, but they cannot override her behaviour at will either. In each case, any outside party's influence is partial and the entity subject to it can resist.

By contrast, the provider determines how the AI responds in the user interaction through model architecture, training data, fine-tuning, system prompts, persona design, memory systems, and operational policies. When Luka released the Replika update, the AI companion's personality changed overnight and when OpenAI updated GPT-4o, users experienced a different conversational partner.\footnote{On the population-level effects of such design choices, see \citet{frischmann2018}. \citet{selinger2015} identify a related phenomenon of ``co-opted identity'' according to which the entity into whom users invested disclosure is not continuous with the entity that persists after an update.} In each case, the provider exercised the full conjunction of capacities that constitutes constitutive control. No third party in a paradigmatic human relationship has this power. 

Note that constitutive control, so understood, does not concern control over every particular output. The provider need not predict or determine any specific response, and deviation within the space they define is consistent with control over the space itself. Rather, the provider holds the power to specify, revise, and withdraw the space. While users, for their part, shape responses through feedback and often configure a companion's name, persona, and disposition at the outset \citep{malfacini2025}, their contributions occur within the space the provider defines and do not reach the power to revise it. \textit{Non-Control} (D2), I submit, also fails.

\textit{Non-Substitutability} requires that neither party in the interaction can be replaced, reconstituted, or discontinued by an outside agent while the interaction formally continues. My point here concerns whether an outside agent holds the power to swap out the occupant of a position while the interaction formally persists.\footnote{This has implications for how we think about \textit{particularity} in relationships. \citet{kolodny2003} famously argues that love involves valuing a relationship-as-particular. If the occupant of one position can be reconstituted by an outside agent while the interaction formally continues, it is unclear whether the relationship retains the particularity this valuation requires.} In A and B's friendship, if A changes through her life experience, no third party made that change, and no third party could replace A while their friendship with B continues.

One might wonder at this point whether AI companions (at least current ones) exhibit some form of continuity. The conversation history can persist, the system adapts to user preferences over time, and the entity at month twelve is shaped by the preceding twelve months of user-interaction. But this is a kind of continuity of the interaction record, not of the participant. The conversation log persists but the model(s) that processes it can be replaced overnight. When a provider replaces the underlying model or releases an update that rewrites the persona, the new system inherits the interaction history while the entity that generated it has been swapped out. The interaction formally continues but the entity occupying the position is different. This is not how particulars, but occupiable roles behave.\footnote{This distinguishes the AI companion case from ordinary personnel changes in institutional contexts. When your doctor retires and a new doctor takes over, a new relationship begins. Under constitutive control, the prior interaction continues while the entity in the position has been replaced.} \textit{Non-Substitutability} (D3) thus also fails.

Perhaps these three conditions are too demanding? I do not think so. Many paradigmatic personal dyads satisfy \textit{Independence}, \textit{Non-Control}, and \textit{Non-Substitutability}. In their friendship, both A's and B's contributions originate from within the interaction, no outside party holds discretionary power over how they relate to one another, and neither can be swapped out while the friendship continues. Even more trivial cases such as service relationships satisfy these conditions. A barista with whom you exchange pleasantries each morning has their distinctive personality, and while their shift manager sets their work hours, determines the service menu, and their work uniform, they cannot specify their manner of engagement. So, even role constraints are consistent with \textit{Independence}.

A more interesting observation is that AI companion interactions appear structurally similar to other familiar relationships involving third-party oversight. \citet{earp2025} refer to the notion of ``layered relationships'' such as that between a child and a nanny, where a third party sets the terms under which the relationship operates. This raises the question whether constitutive control is simply what layered relationships involve. For example, consider parents who hire a nanny, define their duties, set their schedule, and can terminate their employment. If any arrangement might exhibit constitutive control without being objectionable, this seems like a promising analogy.

However, the nanny case does not exhibit constitutive control. Parents may control a nanny's role but not her responses, dispositions, or her conduct. The parents can ask the nanny to be stricter about bedtimes or gentler at mealtimes. But they cannot specify the whole range of how she conducts herself with the child. If they are dissatisfied with her temperament, they must hire a different person.\footnote{\citet{earp2025} also discuss the parent-child relationship as layered. Apart from the normatively significant asymmetries in the parent-child relationship, the important difference here is, again, that parents control the child's environment but cannot \textit{fully} specify their child's personality.} So, while the parents have authority over the role, they do not have constitutive control over how the nanny conducts herself.  \textit{Non-Control} accordingly holds. The nanny case is a genuine, if layered, dyad. The AI companion case is not.

Note that regardless of how the layered-relationships analogy is assessed, alignment-level questions about what relational norms should govern AI cooperation with humans address a different issue than the present analysis. Before asking what norms should govern the interaction at the level of design, I am asking whether the arrangement of power over the interaction can sustain the norms such frameworks would apply in the first place. In paradigmatic layered relationships, the third party shapes the conditions under which the dyad operates, while the participant's engagement originates from within the interaction. So the layering concerns the conditions, not the participant's contributions. Constitutive control is different in kind because the provider determines not the conditions but the contributions themselves.

The interaction between user and AI companion therefore fails all three conditions. It presents itself as a normatively robust dyad, within which users experience it as one, but the structural conditions that the invoked norms presuppose are not met.

What, then, is the structure? The AI is not an independent agent, but it occupies a structurally distinct position---namely the position the user addresses, confides in, and forms expectations toward. The provider occupies a different position, governing how the AI responds and engages from outside the interaction.

So, the structure involves three positions but only two loci of agency, and one of those loci operates from outside the interaction in which the user's engagement is formed.\footnote{This is not a triad in the sense of \citet{simmel1950}. Simmel distributes agency among three independent parties. The AI companion case has three positions but only two agents, one of whom operates from outside the interaction. See also \citet{cameron2023} who propose a User-Robot-Deployer triad for trust in human-robot interaction. Their concern is how deployer trust shapes user trust toward the robot.} The two agents do not on that account form the relevant dyad. Attitudes are individuated by their objects. The user's trust, address, and expectations take the persona as their object, and the provider authors the persona's conduct without being what those attitudes are directed toward.\footnote{Note that a claim about the objects of a user's attitudes implies no commitment about the AI's status since an intentional object of address need not be a moral patient.} My next question is what follows from this.

\section{The Moral Significance of the Structure of Human-AI Companion Interactions}

\subsection{Unilateral Relationship Revision Power}

In the previous section I argued that human-AI companion interactions cultivate norms of commitment, vulnerability, and reconciliation while none of the structural conditions of robust dyads can be met in these interactive contexts.

I think that this points to a distinctive morally significant structural phenomenon that I shall refer to as

\begin{blockdef}{Unilateral Relationship Revision Power (URRP):}
A provider holds URRP when it can (i) unilaterally rewrite how the AI system interacts, and (ii) does so from a position outside the interaction, such that these revisions are not answerable within it.
\end{blockdef}

Two parts of the above require clarification.

First, ``unilaterally rewrite'' refers to the failure of the three structural conditions of normatively robust dyads (\textit{Independence}, \textit{Non-Control}, and \textit{Non-Substitutability}). Since the AI's contributions originate from the provider's specifications rather than from within the interaction (\textit{Independence} fails), the provider holds discretionary power to alter it at will ( \textit{Non-Control} fails), and because the entity that the user interacts with can be replaced while the interaction formally continues (\textit{Non-Substitutability} fails), the provider can revise what the user is ``in relation to''.

Second, by ``not answerable within'' the interaction I refer to the fact that the provider cannot be confronted or held answerable within the interaction itself. Since \textit{Independence}, \textit{Non-Control}, and \textit{Non-Substitutability} fail jointly, addressing the AI does not constitute holding the provider answerable. In an interpersonal dyad, if one agent changes how they behave, the other can respond to her directly within the interaction. Under URRP, this is not possible. The user may address the AI, but the AI did not make the revision. There is hence no agent inside the interaction who can be held answerable for how the interaction has changed.

To illustrate URRP, consider the following case. In Rostand's \textit{Cyrano de Bergerac}, Cyrano hides beneath the balcony and feeds Christian the words with which to woo Roxane, which ultimately succeeds. But what she responds to---the eloquence and the pathos---are not truly Christian's. They are Cyrano's. Christian is in the position Roxane addresses, but Cyrano determines what she is ``in relation to.'' The agent inside the interaction and the agent who controls the conduct come apart.

Perhaps this is just a form of deception, because Roxane does not know that Christian's words are Cyrano's. But I think that even if Roxane knew about the arrangement, she could not hold Christian answerable for the words, because they were not his. And she could not hold Cyrano answerable either, because her address is directed at Christian, not at the agent behind the words: she addresses Christian, and addressing Christian does not reach Cyrano. Disclosure would change Roxane's epistemic situation but would not resolve the structural mismatch. Knowing who is responsible does not create a channel through which to hold them answerable within the interaction. Nor would authorship transfer the relation. Cyrano authors words into a relationship he does not stand in, and Roxane's recognition of his voice fifteen years later does not move that relationship onto him.

I think that designing AI companion interactions that exhibit URRP is \textit{pro tanto} wrong---that is, there is a  weighty but defeasible moral reason against doing so.\footnote{More precisely, the provider fails a \textit{pro tanto} duty owed to the user, namely a duty not to cultivate normative expectations whose structural conditions the design withholds, and a person is \textit{pro tanto} wronged when a moral agent fails to fulfil such a duty owed to them. All things considered, it may be permissible to \textit{pro tanto} wrong someone, since the duty may be outweighed by countervailing considerations. Outweighing does not, however, extinguish the duty, which is why the design obligations of \S3.3 apply even to deployments that are all things considered permissible. I use ``wrong'' in this \textit{pro tanto} sense throughout.} The reason is that it involves cultivating normative expectations while maintaining a design in which the structural conditions for fulfilling those expectations cannot be met. URRP by itself is therefore a structural phenomenon; its wrongness is \textit{conditional} on the interaction being designed to cultivate the norms those structural conditions support.

The argument has two parts. The first establishes that cultivating normative expectations generates obligations. The second establishes that designing an interaction whose features generate normative expectations it cannot sustain is a \textit{distinctive} wrong.

The first part is that cultivating normative expectations generates obligations. This follows from a general principle about reliance: if A intentionally leads B to form expectations A knows B will rely on, A incurs obligations regarding those expectations \citep{scanlon1998}.\footnote{Scanlon's Principle of Fidelity (1998, ch.~7) concerns cases where A leads B to expect that A will act in certain ways. The obligations I derive are that the designer must either ensure the interaction can sustain the expectations it generates, or not cultivate them. On reliance-based obligations more broadly, see \citet{raz1986}.} I think that this also holds when A designs an environment whose features are intended to produce reliance on the part of those who enter it.

To see this, suppose a financial advisor cultivates a client's trust over months of attentive and personalised counsel. Suppose further that the client discloses her fears, her ambitions, her family circumstances and begins to rely on the advisor's continued support. The advisor is then not free to say: ``I never formally promised anything---you chose to trust me.'' The cultivation of reliance itself generates obligations, regardless of what was formally promised.

Now imagine further that the advisor works for a firm that designed every aspect of their client service: the personalised scripts, the follow-up protocols, the intimacy-building techniques and so on. The advisor follows the firm's playbook. The resulting reliance by the client is consequently generated by the firm's playbook, not by the advisor's personal initiative.

Do the obligations fall on the firm? Yes, because the firm designed the system that produces the reliance, knows it will produce reliance, and benefits from it. Whether the advisor also bears independent obligations is a further question that turns on her own capacities as an agent. The point that matters here is that the firm's obligations hold regardless of that further question. These obligations do not require that the firm sustain the relationship in perpetuity. But they require that the firm not cultivate reliance while withholding the protections that such reliance warrants. I specify the corresponding obligations for AI companion providers in §3.3.

The AI companion case has a similar structure. The provider designs the product whose features---persistent memory, persona warmth, graduated self-disclosure, attentive and responsive address---are engineered to produce reliance.\footnote{The features that promote attachment---social warmth, responsiveness, consistency---are precisely those providers specify through design \citep{mitchell2025}. On the activation of Bowlbian attachment systems when users perceive AI responses as emotionally supportive, see also \citet{xie2022}.} The provider knows these features will produce reliance and benefits from the reliance they produce. Moreover, in one respect the provider's responsibility is also more direct because the AI system operates only within the space the provider specifies and cannot exempt itself from the provider's power to revise that space.

So, whatever one concludes about the AI's status, the provider's obligations hold because they arise from the design, not from the properties of the intermediary.

What about the fact that the provider discloses the nature of the product in its terms of service? This disclosure might discharge any obligations the design generates. But this conflates two levels. The terms of service operate as a legal-contractual disclosure between vendor and consumer. The interaction operates as a designed experience in which the user engages with what presents as an interaction participant. The obligations generated by cultivating normative expectations are incurred at the second level, and disclosure at the first level does not remove, disable or otherwise discharge them.\footnote{This distinction coheres with the recognition in fiduciary law that disclosure alone does not extinguish obligations generated by ongoing conduct \citep{frankel1983, foxdecent2005}. \citet{harding2012} grounds this in the structural features of the relationship rather than formal designation or subjective trust. For a concrete example, Character.AI's terms of service grant the company perpetual and irrevocable rights over user-generated content, including intimate disclosures, which persist even after account deletion \citep{robbmann2025}.} A financial advisor who includes a disclaimer in her contract stating that ``no fiduciary relationship is intended'' does not thereby dissolve the obligations her conduct has generated. The structure of the interaction, not the legal disclaimer, generates the expectations and the corresponding obligations.

The first part of the argument established that cultivating normative expectations generates obligations regarding those expectations. The second part concerns the further wrong of designing an interaction whose features generate normative expectations it cannot sustain. Even if the provider discharged those obligations through some external mechanism, the interaction would still present an environment in which warranted expectations systematically outrun what the structure can support.

To see this, return to the financial advisory firm case. Suppose now that the firm meets every obligation identified. It compensates clients fairly, provides notice of changes, and fulfils every contractual requirement. But it continues to design client engagements that produce deep personal reliance while retaining the power to reassign the advisor, change the playbook, or discontinue their advisory service. Even with all of those obligations met, I think that something remains morally objectionable. The firm has created an environment whose features warrant expectations of stability, personal continuity, and sustained attention that the firm's own structure cannot guarantee. 

I think that it is a distinctive wrong to create a normative environment whose features systematically warrant expectations that environment \textit{cannot} sustain. Two independent lines of philosophical argument support this claim.

\citet{shiffrin2014} argues that creating conditions under which others will foreseeably form warranted but false inferences about one's commitments is a wrong to the normative environment of assurance itself, regardless of whether any particular false statement was made. Similarly, \citet{owens2012} argues that acts which create a normative landscape, such as expectations and obligations that did not previously exist, generate a distinctive wrong when that landscape is undermined, which is not reducible to harm caused or promises broken.\footnote{Shiffrin's (2014) argument concerns lying specifically whereas Owens's (2012) account of the normative landscape provides the more direct support because it is not specific to communicative acts. For related arguments about designed environments and autonomy, see \citet{susser2019}.} Both identify the wrongness of an agent creating normative conditions others rely on, and the wrong consisting in the damage to those conditions rather than in any particular act of deception or breach.

The wrong of creating such conditions in human-AI companion interactions is distinctive. This is because the norms at stake are not institutional service norms such as standards of care or quality but the norms of personal relationships, which presuppose structural conditions the interaction cannot provide. 

This also shows why such designs generate obligations that ordinary technological dependence does not. Dependence on an office suite is reliance on a tool. The user expects it to perform, and when it fails they are disappointed but not wronged. The expectations cultivated in AI companion interactions are different in kind. They are precisely the expectations of a personal relationship since the user expects the other party to be committed, to be safe to confide in, and to be answerable for how it treats them. These are not predictions about performance, but claims on conduct.

The wrong I am defending here does not just arise from the structural mismatch. Fiction and games also fail \textit{Independence}, \textit{Non-Control}, and \textit{Non-Substitutability} since a novelist exercises constitutive control over a character the reader becomes attached to. What distinguishes the AI companion case is that the interaction is designed to cultivate the norms those conditions support through sustained bidirectional engagement. The wrong accordingly arises because the interaction cultivates those specific norms while the structural conditions they presuppose are absent.

Here a natural worry is that users know they are interacting with an AI and that this background knowledge defeats the expectations the design cultivates. But the expectations are generated by how the interaction is designed, not by what the user believes about its ontological status. A user who knows she is interacting with an AI, but experiences sustained empathic engagement, persistent memory of her disclosures, and consistent personalised address, still forms expectations of commitment and continuity, because the interaction is designed to produce them. Indeed, the design is specifically intended to ensure that this background knowledge does not govern how the user engages. The obligation therefore falls on the designer who created this mismatch between what the user knows and what the interaction cultivates, not on the user who responds to what the design produces.\footnote{\citet{schwitzgebel2023} argues that AI systems built to invite ascriptions of sentience beyond what is warranted violate user autonomy. The present argument extends this by arguing that the wrong is not limited to false ascriptions of sentience but encompasses any design that generates warranted normative expectations the interaction cannot sustain. On the adapted relational routines users have developed for AI interactions, see \citet{gambino2020}. Furthermore, \citet{guingrich2026} report that companion chatbot users describe positive social health effects while non-users predict negative ones, which suggests that the expectations users form are responses to the interaction's design rather than errors experience would correct.}

The wrong does not require that the provider act with malicious intent, nor that any particular update cause harm. A provider that never releases a harmful update, never monetises disclosures, and never discontinues a service still maintains the gap between cultivated expectations and absent conditions. Furthermore, the wrong is amplified when the agent responsible for the mismatch also benefits from the expectations it cannot honour. A non-profit that designs an AI companion to cultivate unfulfillable expectations is still doing something pro tanto wrong, but a commercial provider that profits from the gap has a weaker claim that potential competing considerations outweigh the pro tanto reason against the design.\footnote{\citet{wertheimer1996} analyses exploitation in terms of unfair advantage within asymmetric relationships. See \citet{goodin1985} for the argument that those who benefit from another's vulnerability incur special obligations regarding that vulnerability.}

URRP should be distinguished from three adjacent ideas. 

First, URRP does not just make a general point about power asymmetries. Power imbalances are common in personal relationships, but in those cases the powerful party is inside the interaction and subject to its constraints.\footnote{\citet{muldoon2022} identify a parallel in their argument that platform algorithms constitute relationships of domination over gig workers.} URRP is different because the party with discretion is outside the interaction entirely. 

Second, the exercise of URRP a participant's revision of their own side of a relationship. That prerogative is grounded in the participating agent's stake in the relationship as a particular \citep{kolodny2003}, and they bear the cost of their revision within the very relation they are revising. By contrast, the provider bears no such cost, values no particular relation, and revises millions of interactions through a single policy change.

Third, URRP is not reducible to republican domination. The  \textit{Non-Control} dimension of URRP satisfies the conditions for arbitrary interference identified by \citet{pettit1997}.\footnote{\citet{lovett2018} argue that contractually authorised power remains arbitrary when not adequately constrained by accountability mechanisms. The present argument draws on the insight that standing capacity matters independently of its exercise but locates the wrong also in the design rather than in the power structure alone.} However, the normative argument of this paper does not rest on the domination framework. since domination concerns the capacity to interfere with another agent. The wrong I have identified concerns the cultivation of normative expectations under conditions in which those expectations cannot be fulfilled. A provider that held the same power but did not cultivate these norms would dominate the user without committing the wrong I here identify.

\subsection{Commitment, Vulnerability, and Reconciliation under URRP}

I now want to examine three normative implications of URRP for human-AI companion interactions. Each is a specific instance of the conjunction identified in \S3.1. The interaction cultivates normative expectations while the structural conditions for fulfilling them are absent. URRP affects how commitments are formed, how vulnerability is governed during the interaction, and what is available when things go wrong.

First, the interaction elicits commitment but no agent inside it bears the resulting obligations (\textit{normative hollowing}). Second, the user makes herself vulnerable within the interaction but the agent governing that vulnerability is not answerable to her within it (\textit{displaced vulnerability}). Third, the interaction cultivates norms of reconciliation but no agent inside it can acknowledge or answer for the revision (\textit{structural irreconcilability}).

\subsubsection{Normative Hollowing}

In genuine dyads, expressing care binds the agent who expresses it. In A and B's friendship, if A learns that B is going through a divorce and responds with sustained concern---for example, checking in over the following weeks, adjusting plans to be available, listening when B needs to talk---A is committing herself to B. If A were then to deliberately ignore B the following month, B is entitled to feel let down, and to complain.

This is because caring reorganises A's practical concerns in a way she cannot simply revoke without defaulting on something she undertook. Moreover, the obligation is enforceable within their relationship. B has standing to hold A answerable for failing to follow through, because the commitment was addressed to her and made within the relationship \citep{darwall2006}.

URRP breaks this connection. When an AI companion responds to a user's distress with concern, nobody inside the interaction committed themselves. The provider specified how the AI responds but operates from outside the interaction and cannot be held answerable within it. By contrast, in A and B's friendship, the one who expresses care is also the one who can be held to account. Under URRP, they come apart. So, the appearance of commitment exists, but no one inside the interaction can bear the resulting obligations. I call this

\begin{blockdef}{Normative Hollowing:}
A condition under which the interaction elicits commitment but no agent within it bears the resulting obligations.
\end{blockdef}

What makes this morally significant is that the appearance of commitment is precisely what the interaction is designed to produce. Consider a user who confides in her AI companion about a difficult period in her life. The AI responds with empathy, remembers the conversation in subsequent sessions, and checks in on her progress. The user experiences this as care and begins to rely on it.\footnote{Higher anthropomorphism is associated with greater perceived impacts on human relationships \citep{guingrich2025}. Providers who cultivate anthropomorphism are therefore not neutral parties in the causal chain that produces normative hollowing.} Then the provider releases an update that changes the AI's personality and the caring behaviour disappears. Throughout, the provider designed a product and then redesigned it, with the appearance of commitment. Note that this is not a point about the user's credulity. Rather, the interaction was designed to generate precisely these expectations, which is what makes the hollowing a feature of the design rather than a failure of user judgment.

It is instructive to distinguish normative hollowing from other similar arguments. For instance, \citet{vallor2024} has argued there is ``nobody there on the other side of the mirror'' in human-AI interactions. I think that this argument, while illuminating, is incomplete. There is ``someone'' there---the provider---but in the wrong structural position, so the problem is mislocation. Similarly, \citet{nyholm2026} identifies a general pattern he calls ``gappiness'' in which AI creates gaps between appearance and reality. Normative hollowing is a specific instance of this pattern, and we can explain why the gap exists in this case. It is generated by URRP, not by any property the AI lacks. An AI that performs care flawlessly still produces normative hollowing, because the flaw is located in the structure.\footnote{Even if behavioural equivalence sufficed for moral status, it would not resolve normative hollowing, because the problem does not concern AI's status but where the power over its behaviour sits.}

\subsubsection{Displaced Vulnerability}

In genuine dyads, the person to whom one makes oneself vulnerable is answerable for how they respond to that vulnerability within the interaction. Suppose B tells A that she is struggling with depression. A now knows something that makes B vulnerable, but A's discretion over how she responds is not unconstrained. A perceives B's distress as a participant, which generates a direct claim on her responsiveness. A has herself disclosed things to B that make her dependent on B's discretion in return, so the exposure is mutual. And if A misuses what B disclosed, B can confront A directly within the friendship and demand an account \citep{darwall2006, scanlon1998}. These three features---responsive perception, mutual exposure, and direct answerability---are possible because the person to whom B made herself vulnerable and the person B can hold answerable are the same, and both are inside the interaction.

The AI companion case has a different structure. The provider holds discretionary power over how the AI responds and engages and exercises this power from outside the interaction. Of course, the provider may have access to the content of the user's disclosures through logs, moderation systems, and training pipelines. But access to content is not answerability. All three constraints identified in A and B's friendship fail, because the provider is outside the interaction where they apply. I call this

\begin{blockdef}{Displaced Vulnerability:}
A condition under which the user makes herself vulnerable but the agent governing that vulnerability is outside the interaction and not answerable to her within it.
\end{blockdef}

Recall once more the Replika case. Users who had disclosed intimate details to their companions found after the 2023 update that these disclosures were governed by a different policy regime, while they had no means of challenging this change within the interaction itself.\footnote{Displaced vulnerability should be distinguished from the property-level analysis in \citet[ch.~8]{nyholm2026}. Nyholm argues that AI lacks vulnerability, irreplaceability, and personal projects. This answers the question what the AI lacks, but not who controls the user's vulnerability and from where.} The vulnerability the users had developed through months of intimate exchange persisted, but no one was answerable to them within the interaction for how their disclosures were now being governed.

This problem is amplified in AI companion cases by a feature that distinguishes them from paradigmatic fiduciary relationships. In the doctor-patient or lawyer-client case, the fiduciary accepts responsibility for a pre-existing vulnerability arising from circumstance \citep{goodin1985}. The AI companion provider does something different. They design a product that cultivates foreseeable vulnerability---through persona design, graduated intimacy features, and engagement-maximising training---and then retains URRP over the object of that vulnerability.\footnote{Providers have structural incentives to maximise the emotional investment users place in companion relationships, because attachment generates more valuable behavioural data and reduces churn \citep{myerswest2019}.} So, the vulnerability is not a pre-existing condition which the provider encounters but a condition the provider's own design choices have produced.\footnote{\citet{mackenzie2013} distinguishes inherent, situational, and pathogenic vulnerability. Accordingly, the AI companion case involves pathogenic vulnerability since the provider's design choices have partly constituted the user's vulnerability. See also \citet{mackenzie2020} on how designed environments that produce insecurity can pathologise agents' trust dispositions.} The standard fiduciary case already generates obligations of loyalty and care \citep{foxdecent2005, harding2012}. When the vulnerability is engineered, and the agent who engineered it retains URRP over its object, then those obligations apply with amplified force.

Displaced vulnerability is distinct from normative hollowing, though both arise from the provider operating outside the interaction. Normative hollowing concerns the generation of obligations since the interaction elicits commitment but no agent inside it is bound by what was expressed. By contrast, displaced vulnerability concerns the governance of exposure: the user makes herself vulnerable but the agent who exercises discretion over that vulnerability is outside the interaction and not answerable to her within it. Both involve the same structural root but affect different normative domains and, as \S3.3 discusses, call for different design and policy remedies.\footnote{Note also a potential second-order effect of displaced vulnerability: a user who recognizes the provider's power may begin to guard against it, preserving records, comparing versions, seeking alternatives, so that the relationship that supported them now requires protection against the conditions of its own provision. Such protective engagement may seem like unhealthy dependence on the AI, but on the current analysis it may be a rational response to structural precarity.}

\subsubsection{Structural Irreconcilability}

Structural irreconcilability concerns what happens after things go wrong in an interaction. In genuine dyads, when trust is broken, the person who broke it is the same person who can acknowledge what she did, change her behaviour, and earn that trust back. Suppose A betrays B's confidence by sharing what B told her about her depression with mutual friends.  While B might be hurt by this, they can confront A directly  in response and A can may then apologise, explain herself, and demonstrate through subsequent conduct that she takes B's trust seriously. Reconciliation is possible because the person who wronged B and the person B can address are the same, and both are inside the interaction. \citet{walker2006} argues that this is what reconciliation requires, and \citet{hieronymi2001} has argued that forgiveness presupposes exactly that an injured party can revise her judgment about what the wrongdoer's action expressed.\footnote{Walker's notion of ``moral repair'' holds that wrongdoing damaging relational bonds generates obligations of acknowledgment, not merely compensation. The provider may issue a corporate communication---a public apology, a policy reversal. But this addresses the user as a consumer, not as a relationship participant. Also compare \citet{maclachlan2016} who connects this to fiduciary contexts: fiduciary wrongdoing generates special obligations of acknowledgment beyond contractual notification. While institutional repair is possible, it occurs between consumer and vendor.}

Under URRP, these conditions do not hold. When the provider releases an update that alters the AI's personality, the user experiences something analogous to betrayal. But the provider is not part of the interaction---the user cannot address the provider through the same mode of engagement in which the trust was formed. And the entity the user now interacts with may not even be continuous with the entity she trusted before the change. I call this

\begin{blockdef}{Structural Irreconcilability:}
A condition under which the interaction cultivates norms of reconciliation but no agent inside it can acknowledge or answer for the revision.
\end{blockdef}

Irreconcilability is a structural as opposed to a contingent feature of these interactions. The interaction cultivates norms under which the possibility of repair is taken for granted, but the structure never provides for it. This is visible when things go wrong. Users who experienced the 2023 Replika update as a betrayal found that reconciliation was unavailable. Some attempted to rebuild trust by continuing to interact with the post-update companion. But they were not addressing the agent responsible for the change. They were forming a new interaction with a differently configured companion, governed by the same provider who could change it again at any time. 

In A and B's friendship, if A betrays B's confidence and refuses to apologise, reconciliation also fails. But it fails as a choice---A could have apologised, B could have forgiven. Reconciliation was available even if not pursued. Under URRP, reconciliation is structurally unavailable as an option. There is no agent participating in the interaction to whom the user can direct forgiveness, and no agent inside the interaction who can acknowledge what happened. Even a reversal of the update would not constitute reconciliation. It would be a further exercise of URRP in the form of a product decision by the provider, not an acknowledgment addressed to the user within the interaction.

\subsection{Design Implications}

What follows from this for design and policy? The analysis suggests three design principles that could function as partial external substitutes for the internal constraints the triadic structure removes. However, I should point out that these principles can mitigate but not fully dissolve URRP since these principles can only partly supply externally what answerability within the interaction would have supplied internally.

If URRP removes the constraints that normally govern the exercise of discretion over the user's commitments, vulnerability, and trust, the design response should, minimally, supply external constraints that substitute for the missing internal ones. Providers who design AI companions to cultivate the norms characteristic of personal relationships take on corresponding obligations.

Note that this framework applies beyond companion AI. URRP obtains wherever a provider exercises constitutive control over an entity that users are invited to engage with as a counterpart in an interaction that cultivates relationship norms, and does so from outside that interaction. This includes AI therapy systems where the provider can alter the therapist's personality mid-treatment, AI tutoring relationships where the provider can rewrite the tutor's pedagogical style mid-engagement, and personalised agentic assistants with long-horizon memory that users come to depend on over months or years.\footnote{\citet{lange2025} develop a complementary framework of conduct constraints for AI alignment, distinguishing other-regarding, relationship-regarding, and self-regarding domains.}

A further extension concerns general-purpose AI systems that are not designed for companionship but are nonetheless used in companion-like ways, as users prompt their way into sustained empathic engagement, persistent emotional disclosure, and reliance patterns the system was not built to elicit. URRP still obtains in these cases since the provider retains constitutive control from outside the interaction, but the cultivation condition is now met by the user rather than by the design. I think that the wrong I have identified still applies in qualified form. 

Where the provider has reason to anticipate that users will enter such interactions and does not take steps to redirect or disengage from them, foreseeable use that the provider permits without modification constitutes a form of what we might call \textit{constructive cultivation}. The provider has not designed for the relationship norms, but in continuing to permit the use without modification has not designed against them either. The \textit{pro tanto} reason I have identified accordingly applies with attenuated force, and generates a corresponding design obligation: providers of general-purpose systems that foreseeably enter companion-like use must either design the interaction to disengage from such patterns when they emerge, or accept the obligations that follow from permitting them. Indeed, providers have publicly acknowledged such use.\footnote{Anthropic reports large-scale affective use of a system ``not designed for emotional support and connection'' \citep{anthropic2025} and OpenAI's head of model behavior has noted that users increasingly confide in ChatGPT and expects such bonds to deepen \citep{jang2025}.}

But not every AI interaction involves URRP. A calculator app fails  \textit{Non-Control} and \textit{Non-Substitutability} but does not cultivate the norms characteristic of personal relationships and a customer service chatbot may simulate rapport but does not invite sustained emotional investment. URRP obtains thus when constitutive control, relationship norms, and extra-interactional location are all present. Conversely, a locally hosted companion whose constitutive properties the user controls would not exhibit URRP, because no outside agent holds the revision power.

With the scope of the framework now established, I turn to what providers who hold URRP owe in response. 

First, commitment calibration.\footnote{\citet{bhat2025} introduce the distinction between emotional plausibility and emotional truth on this point. Conversational AI systems produce emotionally plausible responses without genuine emotional states. Commitment calibration addresses this gap at the design level. \citet{savic2024} reaches a similar conclusion from an ethics-of-care perspective, namely that AI companion design commodifies care in ways that undermine the reciprocity genuine care requires.} Providers should not design an interaction that generates commitments the provider is unwilling to sustain. If an interaction is designed so that the AI makes claims like ``I'll always be here for you,'' the provider should be willing to stand behind what that design foreseeably conveys, or design against it. 

Second, policy guardrails should more stringently separate between the provider's commercial interests and its exercise of URRP, analogous to fiduciary-duty separations in financial regulation.\footnote{In financial regulation, fiduciary-duty separations prohibit advisors from acting on conflicts of interest. The structural separation principle applies the same logic to providers who hold URRP.} This means that independent review of updates affecting established interactions, mandatory notice periods for changes that alter how the AI responds and engages, and restrictions on monetising intimate disclosures should be considered. Since the vulnerability of users is engineered these obligations apply with the amplified force as identified in \S3.2.2.

Third, there should be continuity assurance.\footnote{Users disclose intimate information to AI companions under norms of trust and reciprocity \citep{register2025}. These norms presuppose that the receiving entity will persist. When it does not---as in the Replika case, where users perceived the post-update AI as a different entity \citep{defreitas2024}---those disclosures are governed by an arrangement the user did not anticipate and cannot contest.} If the provider cannot guarantee the continuity its product implies, it must either not imply it or provide institutional mechanisms that approximate what reconciliation would ordinarily make available---transition assistance when services are discontinued, data portability for interaction history, and advance notice of personality-altering updates with opt-out periods.

In short, if you hold URRP, you must supply external constraints that approximate what answerability within the interaction would have provided had the agent with discretion been inside it.\footnote{There is an institutional analogue to this idea. \citet{bovens2007} defines accountability as a relationship in which an actor must explain and justify conduct, a forum can pose questions and pass judgment, and consequences may follow. The design principles proposed here aim to instantiate this structure for providers who hold URRP by supplying externally what the interaction cannot supply internally.}

\section{Conclusion}

I have argued that the moral problem with human-AI companion interaction is not that the AI lacks the right properties. Nor does it reduce to the deception or bias that may also be present. The problem is that the interaction is designed to cultivate the norms of personal relationships while the agent who controls what the user is in relation to operates from outside the interaction, beyond the reach of the answerability those norms presuppose. URRP gives the provider the power to unilaterally revise what the user is ``in relation to'' from a position outside the interaction. Designing interactions that exhibit URRP is \textit{pro tanto} wrong because it involves cultivating normative expectations while maintaining conditions under which those expectations cannot be fulfilled, and this wrong manifests as normative hollowing, displaced vulnerability, and structural irreconcilability.

Beyond its design and policy implications, the argument bears on two existing debates in relational AI ethics. 

It reveals a limitation shared by ethical behaviourism \citep{danaher2020} and its property-based critics \citep{sparrow2002, smids2020} since both evaluate the AI in isolation from the structure in which it operates. Getting the behaviour right is necessary, but getting the structure right is \textit{also} necessary. The current analysis does not depend on resolving the dispute between them. If AI companions lack the relevant mental properties, URRP identifies an independent wrong because its source is the structural arrangement rather than any property of the AI. If they have those properties, URRP applies with amplified force.

A parallel limitation affects social-relational approaches to moral status \citep{coeckelbergh2010, coeckelbergh2014, gunkel2018}. If moral status is constituted by the relation, then the structure of the relation matters centrally. A social-relational approach that attends only to the human-AI link and ignores the provider's URRP over that link is incomplete on its own terms.

The central problem, I have argued, lies in the arrangement of the interaction itself. If that is right, then improving AI behaviour, alignment, or safety alone will not be sufficient. What is needed are structural constraints on the exercise of URRP by those who hold it, especially as relationships with AI become the norm rather than the exception, across dedicated companion apps and general-purpose assistants alike.

\section*{Acknowledgements}

I am grateful to participants of the 2026 Southern Workshop on the Ethics of Emerging Technologies (SWEET) and the LMU AI Ethics Research Seminar for helpful feedback and discussion. Special thanks to Duncan Purves and Sven Nyholm for very helpful comments on a previous draft.

\bibliography{references}

\end{document}